**Surface-confined molecular coolers for cryogenics**

By *Giulia Lorusso, Mark Jenkins, Pablo González-Monje, Ana Arauzo, Javier Sesé, Daniel Ruiz-Molina, Olivier Roubeau,* and *Marco Evangelisti\**

[*]     Dr. G. Lorusso, M. Jenkins, Dr. O. Roubeau, Dr. M. Evangelisti
Instituto de Ciencia de Materiales de Aragón (ICMA) and Departamento de Física de la Materia Condensada
CSIC - Universidad de Zaragoza
C/ Pedro Cerbuna 12, 50009 Zaragoza (Spain)
E-mail: evange@unizar.es
Homepage: http://molchip.unizar.es/

P. González-Monje, Dr. D. Ruiz-Molina
Centre d'Investigació en Nanociencia i Nanotecnologia (CIN2, CSIC)
Esfera UAB, Edifici CM7
Campus UAB, 08193 Cerdanyola del Vallès (Spain)

Dr. A. Arauzo
Servicio de Medidas Físicas
Universidad de Zaragoza
C/ Pedro Cerbuna 12, 50009 Zaragoza (Spain)

Dr. J. Sesé
Instituto de Nanociencia de Aragón (INA) and Departamento de Física de la Materia Condensada
Universidad de Zaragoza
C/ Mariano Esquillor s/n, 50018 Zaragoza (Spain)

In the search for smaller, faster, more selective and efficient products and processes, the engineering of spatial nano- and micro-arrangements of pure and composite materials is of vital importance for the creation of new devices. A very representative example of versatility and potentiality arises from the field of molecular magnetism since it provides a privileged way to synthesize magnetic nanomaterials with a variety of physical properties, in macroscopic amounts and of homogenous size.[1] Exploiting the functionality of, so-called, molecular nanomagnets has led to their potential use as magnetic refrigerants for liquid-helium temperatures.[2] At the basis is the Magneto-Caloric Effect (MCE), according to which the cooling proceeds following the removal of an applied magnetic field under adiabatic conditions.[3] By chemically engineering the molecules as such to optimize characteristics like magnetic anisotropy, type of exchange interactions and metal:non-metal ratio, the MCE can be notably enhanced to much larger values than that encountered for conventional magnetic refrigerants made of lanthanide-alloys or magnetic nanoparticles.[2] Concurrently, research on surface-deposited molecular aggregates has been evolving with the aim of assembling and integrating molecules into on-chip functional devices.[4] In this regard, sub-kelvin microrefrigeration will allow reducing large quantities of refrigerants and simplifying the use of sophisticated equipment. These mesoscopic devices will then find application as cooling platforms for all those instruments where local refrigeration down to very low temperatures is needed, such as high-resolution X-ray and gamma-ray detectors for astronomy, materials science, and security instrumentation. Furthermore, this technique could open new markets by making available cheap ($^3$He-free) cooling. The idea of employing molecular nanomagnets for this purpose is innovative and markedly in contrast with the electronic schemes which are currently explored.[5] Obviously for this approach to become a reality, a relatively strong binding of the molecules to the surface and the preservation of their functionalities once deposited are sine-qua-non conditions.

The magnetothermal investigations on molecular nanomagnets have so far been carried out exclusively for bulk materials, whereas their transposition to surfaces is challenging, both for the low temperatures required and, specially, for the very weak strength of the magnetic signal arising from the surface. In this Communication we focus on [$Gd_2(CH_3COO)_6(H_2O)_4$]·$4H_2O$, hereafter shortened as $Gd_2$-ac (see **Figure 1**), i.e., a previously studied ferromagnetic molecular dimer showing one of the largest MCEs reported to date for liquid-helium temperatures.[6] As a step towards the interfacing of this molecular nanomagnet with Si-based thermal sensors designed to function as microrefrigerators, and to address their magnetothermal properties at the nanometric scale, we selectively deposit Gd$_2$-ac molecules on a silicon surface via a tip-assisted technique, i.e., Dip-Pen Nanolithography (DPN). A detailed investigation of the magnetic stray field generated by the as-deposited molecules was then carried out by Magnetic Force Microscopy (MFM) near liquid-helium temperatures and in moderate/high applied fields. We shall see below that a quantitative analysis of the MFM images permits us to conclude that the molecules hold intact their magnetic properties, and therefore their MCE and cooling functionality, after their deposition on the Si substrate.

Previous to surface magnetic measurements, structuration of Gd$_2$-ac molecules on Si substrate is needed to ensure a proper contrast between magnetic and non-magnetic areas as needed to estimate the magnetic stray field generated by the deposits. For this, DPN is a suitable technique since it has already been shown to precisely place drops of a controlled size according to predefined patterns with sub-micrometer precision.[7] For the substrate we make use of Si wafers that are p-doped with boron to improve its conductivity ($\rho \approx 0.1$ $\Omega$/cm) and to permit its grounding, particularly important for preventing the accumulation of electric charges during MFM measurements. Furthermore, we pattern a (75 x 75) μm$^2$ grid by means of Focused Ion Beam (FIB) to help locate the final molecular arrays for the AFM/MFM experiments. As a last step before the deposition of the molecular nanomagnets, we clean the wafers using ultrasound in acetonitrile, ethanol and deionized water to provide a clean writing surface. Besides providing a clean surface, this last step also ensures the presence of a thin layer of native oxide,[8] which in turns enables the adsorption of molecular species through hydrogen bonding with hydroxyl groups naturally present at the surface of oxides, even without specific pre-treatment.[9] With its four terminal coordinated water molecules and acetate groups in various coordination modes, the neutral Gd$_2$-ac molecule may form a range of hydrogen bonds, either as donor or acceptor, with surface hydroxyls or adsorbed water, as it indeed does in its crystalline form with lattice water molecules (**Figure 1**).[6] Gd$_2$-ac is thus a good candidate to be efficiently attached to hydrophilic surfaces without pre-functionalization, although with no control over the orientation of the molecule.

The ink used for the $Gd_2$-ac deposits consists of a 5 mg/ml solution of $Gd_2$-ac in a mixture of dimethylformamide (DMF) and glycerol, at 95% and 5% by volume, respectively. To support that the molecules in solution, and most importantly in the final deposits, preserve a similar structure as those in bulk $Gd_2$-ac, we follow through Attenuated Total Reflectance (ATR) Infra-Red (IR) spectroscopy the evaporation of a macroscopic drop of the exact same solution, until it forms a sticky white thin film on the ATR crystal. While the IR spectra of concentrated solutions only show hints of bands of $Gd_2$-ac, the spectra of the final sticky film match well that of crystalline $Gd_2$-ac plus bands due to traces of DMF and glycerol (**Figure S1**). Because slight modifications of the bridging geometry of the acetate ion would not result in significantly different spectra, this does not imply that deposited molecules have the exact same structure as in bulk $Gd_2$-ac, but it does corroborate our initial assumption. Direct spectroscopic characterization of the deposits is not feasible because of the limited amount of material. Under these experimental conditions and controlling the temperature and humidity, maintained constant at 25ºC and 40%, respectively, we make use of DPN to obtain structures which are reproducible and uniform in size. Among possible patterning geometries there are circles with a diameter of up to 30 μm, a representative portion of which is in the in-air AFM image of **Figure 2.a**, which shows a (11 x 11) $μm^2$ area of the deposited sample. It can be seen that the molecular aggregates form slightly oval-shaped drops, the elongation being caused by a drift of the AFM tip during DPN deposition. A mean drop size can be estimated from the profile reported in **Figure 2.b**, which provides ≈ 10 nm height, while the length of the two oval axes is ≈ 1.7 and 1.4 μm, respectively. Molecules of $Gd_2$-ac in such multilayer deposits will likely form hydrogen bonds among them, in a similar manner as in their crystalline state, resulting in a dense packing within each drop. Therefore, from the estimate of the drop size and from the density of $Gd_2$-ac, we estimate that each single drop should have a maximum magnetization ≈ 2 x $10^8$ $\mu_B$ at the saturation (see Supporting Information and **Figure S2** for further details).

Next, we present the low-temperature MFM experiments which are performed by focusing our microscope on a single representative $Gd_2$-ac drop. We here remind that this molecular dinuclear complex is characterized by an intramolecular ferromagnetic exchange interaction of relatively weak strength ($J/k_B$ < 0.07 K for a Hamiltonian of type $H = -J$ **S**·**S**).[6] We will safely neglect this coupling in analyzing the in-field MFM images below, since the latter are collected for fields $|B| \geq 0.5$ T, i.e., sufficiently large to magnetically decouple the $Gd^{3+}$ spins. In **Figure 3** we report MFM images collected in the frequency shift ($\Delta f$) mode, together with the corresponding profiles for different applied fields at $T$ = 5 K. For comparison, in **Figure S3** we report a similar set of data, though collected at $T$ = 9 K. All images are taken on the same scan area of (2.3 x 2.3) $\mu m^2$, with a resolution of 500 lines. The amplitude of the cantilever oscillation is 10 nm, while the tip resonance frequency is $f_0 \approx 71$ kHz for all images. For $B$ = 0, we expect no magnetic stray field from the $Gd_2$-ac drop. Therefore in order to minimize the van der Waals contribution, we set the tip-to-sample distance as such to barely see any topography for zero-applied field, which we accomplish when the tip is higher than $h \approx 150$ nm. The area external to the drop is the non-magnetic contribution of the substrate which constitutes our reference background, $\Delta f_{bg}$ (dashed lines in **Figure 3**). Before collecting each image, we withdraw the tip to a safe distance, set the new applied magnetic field, retune the resonance frequency, and finally approach the tip until we meet the condition $\Delta f = \Delta f_{bg}$ on top of the non-magnetic area. This procedure guarantees that all images are taken at the same $h$. In addition, electrostatic interactions between the tip and the sample are compensated by a bias voltage of 430 mV.

All in-field MFM images are collected for applied magnetic fields largely exceeding the coercive field of the tip ($\approx$ 500 Oe). Therefore, the magnetization of the tip constantly is at its maximum value, $M_t$, during the time of each measurement. Confirmation of this assumption is obtained by collecting a MFM image for a reverse field $B = -1$ T, to be compared with its *identical* counterpart for $B = 1$ T (**Figure S4**). Indeed, this field strength is sufficient to flip and polarize both tip and sample magnetizations; therefore the tip-sample interaction does not change by inverting the applied field.

The evolution of magnetic contrast between the $Gd_2$-ac drop and the non-magnetic substrate is well visible in **Figure 3**, as a function of the applied magnetic field. Specifically, the inner area of the drop becomes darker, while the border brighter and thicker, by increasing the field. The bright border is an evidence of the inversion of the stray-field flux lines from $Gd_2$-ac in proximity of the border of the drop. The profile lines reported in **Figure 3** for each corresponding MFM image provide further evidence of the dependence of the magnetic contrast on the applied field. All profiles refer to the straight line vertically bisecting the drop and are obtained by making use of WSXM software analysis.[10] Following the same procedure, we perform MFM measurements vs. $B$ at $T = 9$ K, and the results are reported in **Figure S3**. An analogous trend is nicely visible.

In order to quantitatively analyze the collected MFM images, we consider the dependence on $B$ and $T$ of the maximum frequency shift, $-\Delta f_{max}$, i.e., the height of the profiles in **Figures 3** and **S3**. We first note that the relatively large and varying applied field induces fluctuations in the sample magnetization solely, while no other experimental parameter is perturbed, viz., the magnetization tip is constantly saturated. Therefore in this limit of independent sample and tip, the frequency shift, which measures the gradient of the force acting on the tip, has to be directly proportional to the stray field generated by the drop, thus to the $Gd_2$-ac magnetization. We denote $-\Delta f_{max} = M \cdot c^{-1}$, where $M$ is the $Gd_2$-ac molar magnetization and $c$ is a proportionality constant – a similar approach has recently been used for the direct measurement of the magnetic moment of individual nanoparticles.[11] In **Figure 4**, $-\Delta f_{max} (T,B)$ is then compared with the isothermal magnetization curves of $Gd_2$-ac, as obtained both from magnetization experiments on a massive bulk sample (empty circles) and from calculating it as the sum of two paramagnetic Gd$^{3+}$ ($s = 7/2$, $g = 2.0$) spin centers (solid lines), at the corresponding temperatures. The ($T,B$)-dependence of the MFM signal beautifully follows the same trend of the isothermal magnetization curves of Gd$_2$-ac bulk material, providing $c \approx 7\ N\mu_B\mathrm{Hz}^{-1}$, and undoubtedly demonstrating that the magnetic properties of Gd$_2$-ac are preserved in the deposited drops.

To further facilitate the interpretation of our experimental results we elaborate a model within the point dipole approximation,[12] according to which the tip is reduced to a magnetic dipole – see Supporting Information for full details. The tip is let to interact with the Gd$_2$-ac molecules positioned within each drop via dipolar interactions. The $\Delta f$ is then computed for the same applied fields and temperatures we employ in our experiments. **Figures 3**, **4** and **S3** show the so-obtained simulations which nicely compare with the behavior experimentally observed. In addition to the verification of the anticipated paramagnetic-compatible dependence of $-\Delta f_{max}$ (**Figure 4**), one can notice that the units of all simulated curves scale with the corresponding experimental values by a factor which remarkably is well below one order of magnitude. Even though this factor is determined by multiple parameters (e.g., encompassing the tip size, shape, height and magnetic moment, and similarly the sample position and magnetization), we stress that these parameters' values are set in close agreement with our experiments – see Supporting Information.

Finally, the sensitivity of our MFM measurements is determined according to the following procedure. The sample magnetization at saturation ($\approx 2 \times 10^8\ \mu_B$ for an individual drop – see S. I.) corresponds to a detected $\Delta f = 2.0$ Hz (**Figure 3**). On the other end, frequency shifts below $\approx 0.25$ Hz (for $B = 0.5$ T and $T = 5$ K, in **Figure 3**) are hardly detectable in our experimental conditions because of the thermal noise on the cantilever. Therefore for $h \approx 150$ nm, we obtain the MFM sensitivity ≈ 0.25 / 2.0 · 2 x $10^8$ $\mu_B$ ≈ 2.5 x $10^7$ $\mu_B$, corresponding to ≈ 3 x $10^{-16}$ A $m^2$.

To summarize, MFM is used near liquid-helium temperature and up to $B$ = 9 T for measuring the stray field generated by $Gd_2$-ac molecular aggregates in the form of drops, deposited on Si surface. The ($T$,$B$)-dependence of the stray field measurements is akin to that of the magnetization of the bulk equivalent magnetocaloric material, thus enabling us to conclude that the as-deposited $Gd_2$-ac molecules hold intact their magnetic characteristics. We finally note that the collective behavior found in the bulk equivalent material, i.e., a magnetic phase transition at $T$ ≈ 0.2 K driven by dipolar interactions,[6] should likely be affected by the reduced thickness of the drops (≈ 10 nm), favorably pushing the magnetic ordering to even lower temperatures. The lowest temperature which can be attained in a process of adiabatic demagnetization should therefore be lowered likewise.[13] Transferring a known, excellent cryogenic magnetocaloric material, such as the $Gd_2$-ac molecular nanomagnet, from bulk crystal to Si substrate without deterioration of its properties, paves the way towards the realization of a molecule-based microrefrigerating device for very low temperatures.

**Experimental Section**

*Material.* All commercial reagents and solvents are of analytical grade and used without further purification. $Gd_2$-ac is synthesized as described previously[6,14] and initially obtained as single-crystals. Purity is checked by single-crystal and powder X-ray diffraction. Solutions are made by dissolving a powdered bulk sample in a mixture of dimethylformamide (DMF) and glycerol, at 95% and 5% by volume.

*Substrate.* We employ polished (100)-oriented Si wafers with boron doping (type p). The relatively low electrical resistivity ($\rho$ ≈ 0.1 Ω/cm) assures good grounding. To help locate the molecular aggregates, we pattern (75 x 75) µm² grids of indexed trenches etched with a Dual-Beam (SEM/FIB) Helios 600 by FEI.

*DPN*. DPN experiments are performed with a NScryptor DPN System (from NanoInk, Inc.). All DPN patterning processes are carried out under constant conditions, room temperature and ~40% of relative humidity, using an integrated environmental chamber as part of the NScryptor DPN System. Commercially silicon nitride Type A Single pens, with a spring constant of 0.1 N·m$^{-1}$, are used in all DPN experiments. Tips were coated using a microfluidic ink delivery chip-based system (Inkwell, from NanoInk, Inc.). The inkwells contain several reservoirs that are filled with the desired solution and transferred to the microwells. Here, the tip is coated with the Gd$_2$-ac solution by dipping. Gd$_2$-ac nanoarrays are generated by traversing the tip over the surface in the form of the desired pattern, after removing the excess of material from the tip in order to achieve uniform dots.

*AFM*. We use a Ntegra Aura AFM by NT-MDT working in air, at room temperature and in semi-contact mode.

*MFM*. We use a Nanoscan high-resolution cryo-AFM/MFM for variable magnetic fields, which can be operated in a PPMS Quantum Design. The high-resolution MFM images are collected in non-contact mode. The recorded magnetic contrasts result from the change in the frequency resonance of the cantilever: an attractive tip-sample interaction, increasing on sample approaching, shifts the resonance to lower frequency (darker), while a repulsive interaction shifts the resonance to higher frequency (brighter).[15] We employ a high-resolution MFM tip by Team Nanotec GmbH with Co alloy coating, radius < 25 nm, spring constant $k \approx 0.7$ N/m, resonance frequency $f_0 \approx 71$ kHz and coercive field $\approx 500$ Oe. Before collecting the zero-applied-field image, the tip was pre-magnetized along the tip axis, normal to the sample.

**Acknowledgments**

We thank Dr. T. V. Ashworth and Prof. E. K. Brechin for fruitful discussions. This work has been supported by the Spanish MINECO through grants MAT2009-13977-C03, MAT2011-24284 and CSD2007-00010, by a CSIC JAE-technician fellowship (to P. G.-M.) and by an EU Marie Curie IEF (to G. L.).

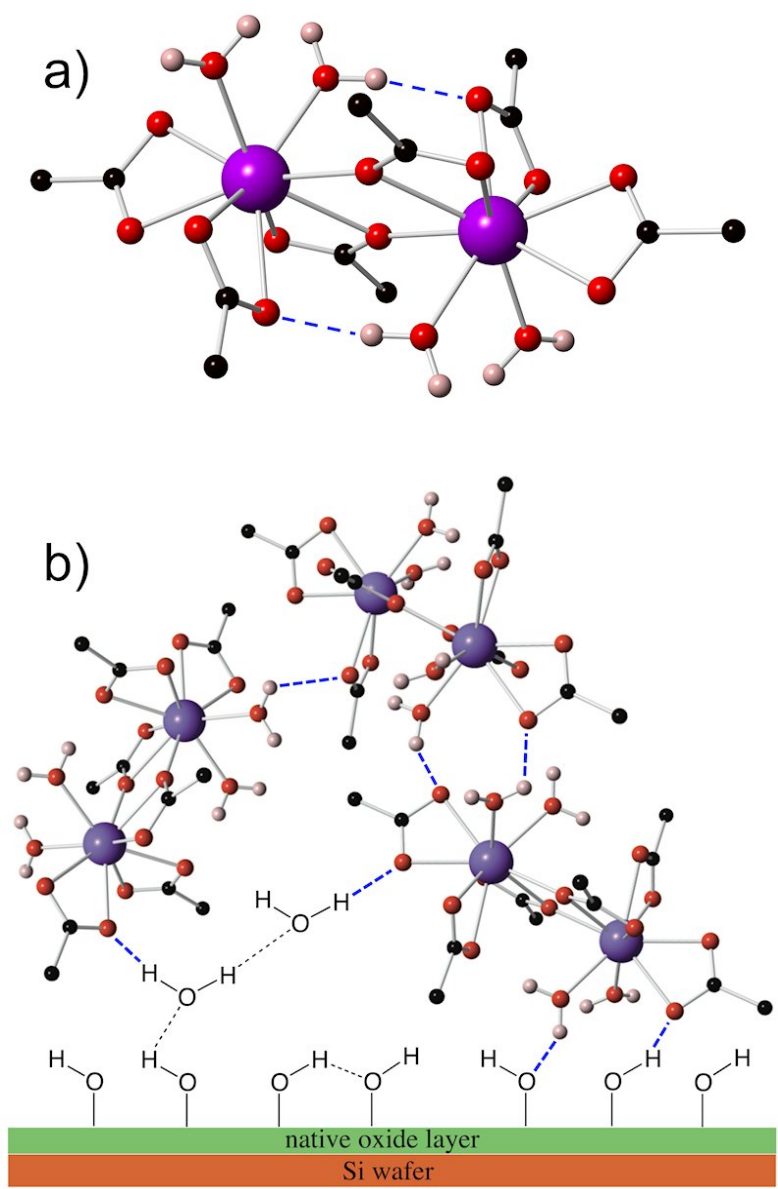

**Figure 1.** a) Molecular structure of the dinuclear neutral complex in Gd$_2$-ac. Dashed blue lines highlight the intramolecular hydrogen bonds, increasing the stability of the molecule. b) Schematic hypothetical representation of Gd$_2$-ac deposited on a Si wafer showing some of the many possible interaction paths through hydrogen bonding involving the surface silanol groups, adsorbed water and the Gd$_2$-ac water and carboxylic groups.

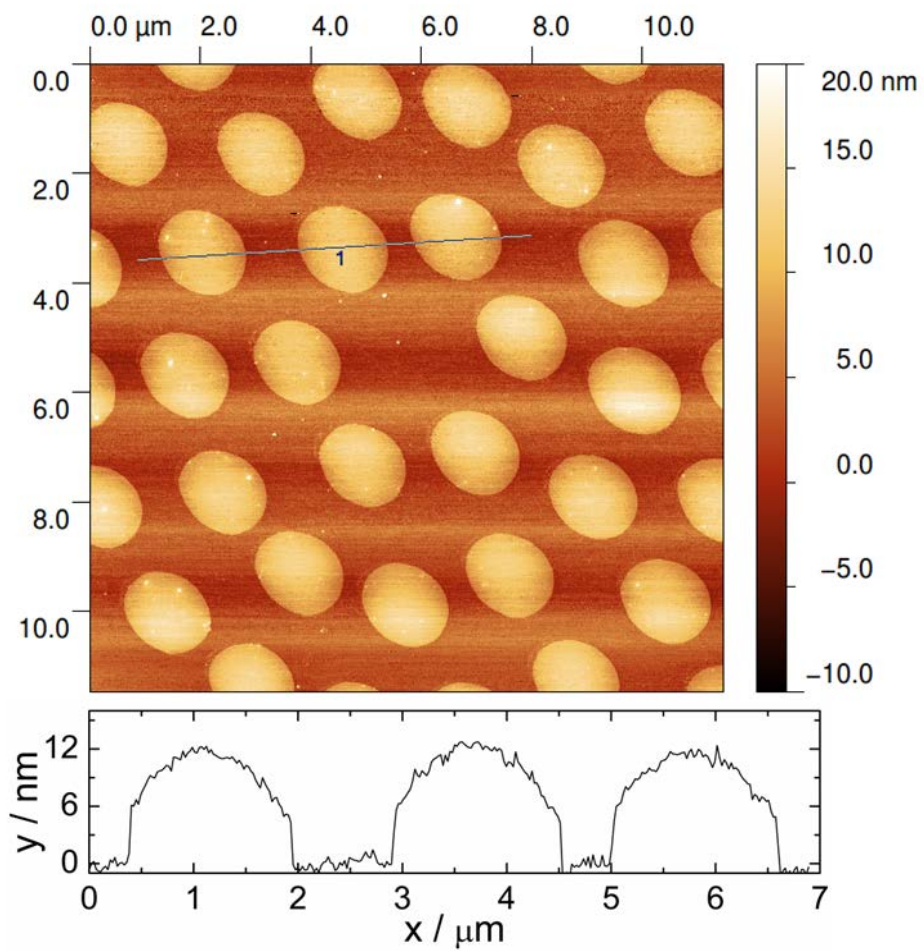

**Figure 2.** Room-temperature topography AFM image of the $Gd_2$-ac drops deposited on silicon wafer by DPN. Height and width of the drops are obtained from the profile relative to the straight line 1, reported in the bottom panel.

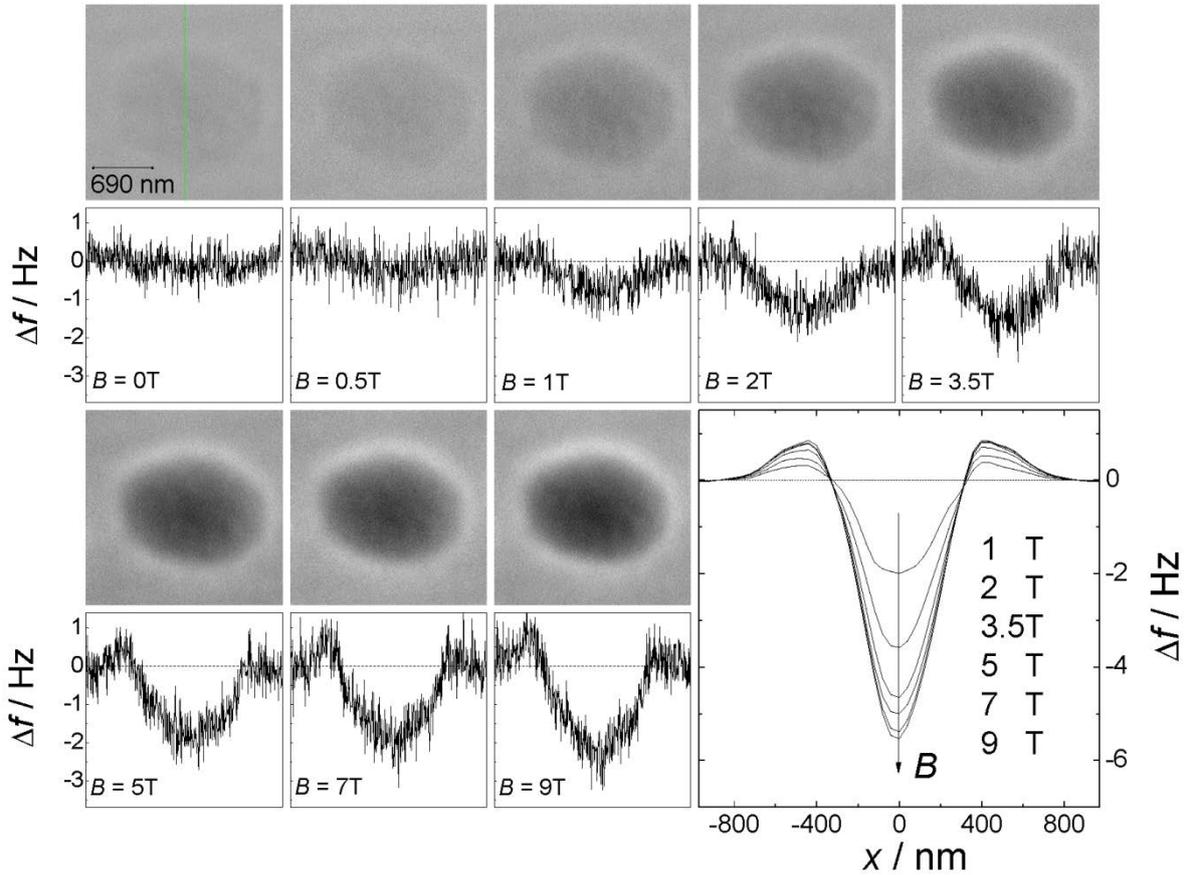

**Figure 3.** MFM frequency shift, $\Delta f$, images of a single Gd$_2$-ac drop taken at different magnetic fields, as labeled, and $T = 5$ K. The images are represented in the same contrast scale, namely $-3.4 \div 1.5$ Hz. Magnetic profiles are presented below each corresponding image, with the background level -see text- being represented by a dashed line. Bottom-right panel is the simulated $\Delta f$ within the point dipole model for $T = 5$ K and selected magnetic fields, as labeled.

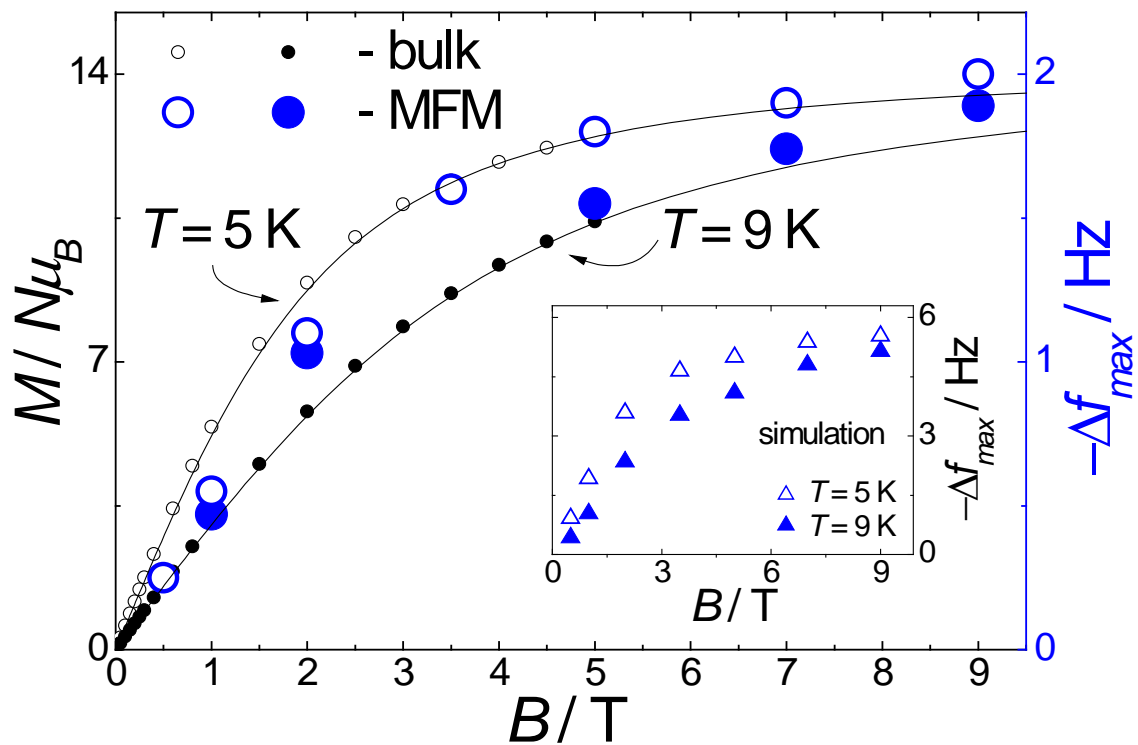

**Figure 4.** Maximum frequency shift, $-\Delta f_{max}$ ($T,B$), obtained from **Figures 3** and **S3**, for as-deposited Gd$_2$-ac, together with experimental[6] and calculated (solid lines) isothermal magnetization curves for the bulk equivalent material, as labeled. Inset: Calculation of $-\Delta f_{max}$ within the point dipole model.

**Table of contents entry.**

*Abstract:* An excellent molecule-based cryogenic magnetic refrigerant, gadolinium acetate tetrahydrate, is here used to decorate selected portions of silicon substrate. By quantitative magnetic force microscopy for variable applied magnetic field near liquid-helium temperature, we demonstrate that the molecules hold intact their magnetic properties, and therefore their cooling functionality, after their deposition. Our result represents a step forward towards the realization of a molecule-based microrefrigerating device for very low temperatures.

*Authors:* Giulia Lorusso, Mark Jenkins, Pablo González-Monje, Ana Arauzo, Javier Sesé, Daniel Ruiz-Molina, Olivier Roubeau, and Marco Evangelisti*

*Title:* Surface-confined molecular coolers for cryogenics

*Figure:*

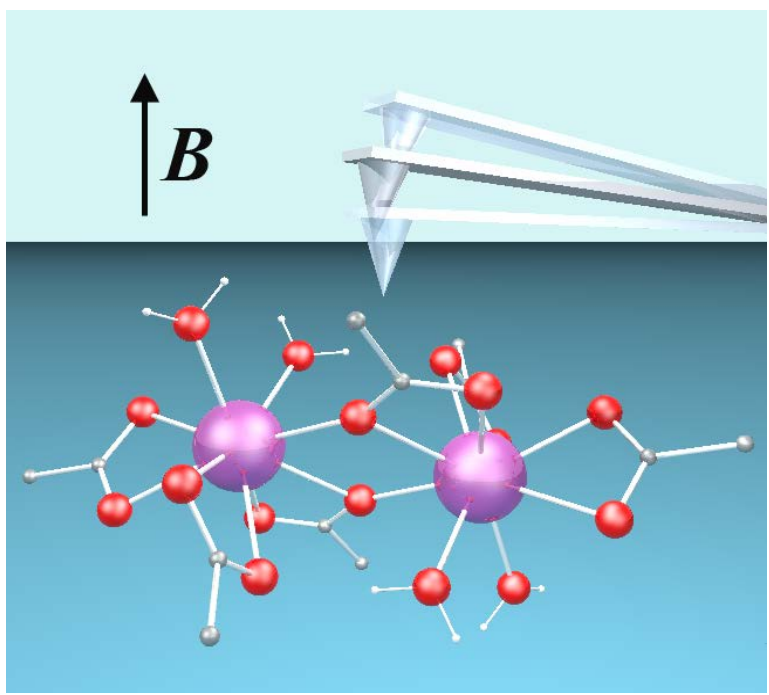